\documentclass{article}

\usepackage{arxiv}

\usepackage[utf8]{inputenc} 
\usepackage[T1]{fontenc}    
\usepackage{hyperref}       
\usepackage{url}            
\usepackage{booktabs}       
\usepackage{amsfonts}       
\usepackage{nicefrac}       
\usepackage{microtype}      
\usepackage{lipsum}		
\usepackage{graphicx}
\usepackage{doi}
\usepackage{multirow, multicol}
\usepackage[section]{placeins}
\usepackage[square,sort,comma,numbers]{natbib}

\title{Comparison of methods for analyzing environmental mixtures effects on survival outcomes and application to a population-based cohort study}

\date{}	

\author{
  Melanie N. Mayer\thanks{Corresponding author: mm4963@cumc.columbia.edu} \\
  Department of Biostatistics\\
  Columbia University\\
  New York, NY 10032 \\
  \texttt{mm4963@cumc.columbia.edu} \\
   \And
  Arce Domingo-Relloso \\
	Department of Biostatistics\\
	Columbia University\\
	New York, NY 10032 \\
	\texttt{ad3531@cumc.columbia.edu} 
 \AND
 Marianthi-Anna Kioumourtzoglou \\
	Department of Environmental Health Sciences\\
	Columbia University\\
	New York, NY 10032 \\
	\texttt{mk3961@cumc.columbia.edu} \\
  	\And
Ana Navas-Acien \\
	Department of Environmental Health Sciences\\
	Columbia University\\
	New York, NY 10032 \\
	\texttt{an2737@cumc.columbia.edu} 
   	\And
    Brent Coull \\
	Department of Biostatistics\\
	Harvard University\\
	Boston, MA 02115 \\
	\texttt{bcoull@hsph.harvard.edu} 
   	\And
    Linda Valeri \\
	Department of Biostatistics\\
	Columbia University\\
	New York, NY 10032 \\
	\texttt{lv2424@cumc.columbia.edu} 
}

 \renewcommand{\headeright}{}
\renewcommand{\undertitle}{}
\renewcommand{\shorttitle}{\textit{} }

\begin{document}
\maketitle

\begin{abstract}
The estimation of the effect of environmental exposures and overall mixtures on a time-to-event outcome is common in environmental epidemiological studies. While advanced statistical methods are increasingly being used for mixture analyses, their applicability and performance for survival outcomes has yet to be explored. We identified readily available methods for analyzing an environmental mixture’s effect on a survival outcome and assessed their performance via simulations replicating various real-life scenarios. Using prespecified criteria, we selected Bayesian Additive Regression Trees (BART), Cox Elastic Net, Cox Proportional Hazards (PH) with and without penalized splines, Gaussian Process Regression (GPR) and Multivariate Adaptive Regression Splines (MARS) to compare the bias and efficiency produced when estimating individual exposure, overall mixture, and interaction effects on a survival outcome. We illustrate the selected methods in a real-world data application. We estimated the effects of arsenic, cadmium, molybdenum, selenium, tungsten, and zinc on incidence of cardiovascular disease in American Indians using data from the Strong Heart Study (SHS). In the simulation study, there was a consistent bias-variance trade off. The more flexible models (BART, GPR and MARS) were found to be most advantageous in the presence of nonproportional hazards, where the Cox models often did not capture the true effects due to their higher bias and lower variance. In the SHS, estimates of the effect of selenium and the overall mixture indicated negative effects, but the magnitudes of the estimated effects varied across methods. In practice, we recommend evaluating if findings are consistent across methods.
\end{abstract}


\section{Introduction}

Environmental mixtures studies often consist of multiple continuous and correlated exposures with complex relationships, such as nonlinear and interaction effects, with the outcome.\cite{Billionnet2012, Carlin2013} Historically, it has been common to consider the effect of a single exposure at a time; however, recently there has been a push to consider the complete mixture to better capture what occurs in real life. \cite{Carlin2013, Dominici2010, Coull2015, Taylor2016} Moreover, environmental epidemiological studies are often concerned with the impact of exposures on the time of development of a particular outcome, such as death or disease.\cite{Dockery1993, Domingo2019, Smarr2021} For example, the Strong Heart Study (SHS) is a cohort study that aims to assess the risk factors associated with incidence of cardiovascular disease (CVD) in American Indians. \cite{Lee1990} To analyze time-to-event outcomes (also known as survival outcomes), specialized advanced methods are required.

Given the complexities associated with modeling environmental mixtures in health studies, various statistical methods have been developed specifically for this context. Recent studies have discussed and compared the performance of existing mixture methods when interest focuses on modeling an environmental mixture's effect on a health outcome. \cite{Lazarevic2020, Gibson2019, Hamra2018} However, fewer mixtures methods exist for survival settings, and a comparison of the performance of the few existing methods applicable to survival outcomes has not yet been conducted. \cite{Joubert2022} There is no assurance that a method's performance when modeling a survival outcome is comparable to that when modeling a continuous or binary outcome. While we don’t expect there to be a single best-performing method under every scenario given variations in the research question, the data structure, and the true exposure-response curve form, insight into their applicability and performance is helpful. Without assessing/evaluating them in the survival context, we run the risk of applying a sub-optimal method to estimate effects of an environmental mixture, potentially resulting in non-reproducible findings. Additionally, with the recent increase in machine learning (ML) approaches available, it may be beneficial for environmental epidemiologists to utilize these flexible modern techniques on studies of environmental mixtures with survival outcomes to better capture the complex relationships and achieve accurate conclusions. 

Our main interest was to compare the bias and efficiency of mixtures’ exposure effect estimates from traditional and ML models for survival outcomes. Although many different research questions about mixtures exist, we focused on estimating individual exposure, overall mixture, and interaction effects on a survival outcome in the context of non-time-varying exposures. To do so, we offer an extensive literature search and understanding of the operating characteristics of currently available survival time model approaches in the context of environmental mixtures. Many ML methods have been extended to the survival outcome setting.\cite{Biganzoli2002, Ishwaran2008, Shivaswamy2007} However, not all are applicable due to limitations such as inability to draw inferences or excessive runtimes. After filtering through available methods, we considered Bayesian Additive Regression Trees \cite{Sparapani2016}, Multivariate Adaptive Regression Splines \cite{LeBlanc1999}, Cox Elastic Net \cite{Simon2011}, and Gaussian Process Regression \cite{Williams1998}. Additionally, we compared these ML approaches to the Cox Proportional Hazards model including all mixture members simultaneously, with and without penalized splines, due to its popularity in environmental health sciences. \cite{Cox1972, Malloy2017}

We performed a simulation study to evaluate the performance of the selected methods under different real-world scenarios, including ones involving nonlinear effects and proportional hazards violations. Once we better understood the strengths and limitations of the methods, we applied them to data from the SHS to estimate the association between exposure to multiple metals and incident CVD. We leveraged our current knowledge on the association of metals with incident CVD in the SHS to motivate relevant research questions, compare how well the estimates align with past results, and extend our knowledge by also considering the overall mixture effect on incident CVD. Most analyses of the SHS have focused on the assessment of one or two metals at a time. \cite{Moon2013, Nigra2018, Tellez2013} In this analysis we included six metals (arsenic [As], cadmium [Cd], molybdenum [Mo], selenium [Se], tungsten [W], and zinc [Zn]). By evaluating the performance of various methods for studying environmental mixtures' effect on survival outcomes, we aim to guide modeling choices in future studies, illuminate our current capabilities to achieve this objective, and support future methodological development.

\section{Methods}\label{sec2}
\subsection{Estimands of Interest}

We considered the scenario where we have baseline, time-invariant measures on multiple environmental exposures and confounders. Subjects are followed through a follow-up period where the occurrence of the event of interest is monitored. We are interested in estimating the effect of (1) an individual mixture component, (2) the overall mixture, and (3) interactions among mixture components on a survival outcome. To quantify each of these three types of effects, we considered popular estimands for time-to-event outcomes. The hazard function, $\lambda(t)= \lim\limits_{\Delta t \to 0}\frac{Pr(t \leq T \leq t+ \Delta t|T>t)}{\Delta t}$, is the instantaneous probability of experiencing an event at time $t$ given that a person has not experienced an event before time $t$. The survival function, $S(t)= P(T > t)$, is the probability of surviving beyond time $t$. Traditionally, the Cox Proportional Hazards (PH) model, which estimates hazard ratios (HRs), has been used to model survival outcomes.\cite{Cox1972} Thus, a natural estimand used by environmental epidemiologists has been the HR, which quantifies changes in hazards at different levels of the exposure of interest on the multiplicative scale. To be more comprehensive, we also considered effects with respect to the survival probability difference scale. Different estimands may be of interest for public health decision-making, and it has been shown that conclusions can vary depending on the choice of scale being used (i.e. ratio vs difference measures). \cite{Spiegelman2017} Furthermore, when it comes to survival analysis, the clinical meaning of a HR can be difficult to interpret, especially when the underlying proportional hazards assumption is violated. A model-free estimand with clearer clinical and analytic interpretability, such as the survival probability difference, may be preferred in certain circumstances.\cite{Uno2014}

To estimate the effect of exposures $j = 1, ....J$ on a survival outcome, we use $\lambda^{k_1,...,k_{|j|}}_{\mathbf{j}\subset \{1,...,J\}}(t|\mathbf{M^{(-j)},C})$ and $S^{k_1,...,k_{|j|}}_{j\subset \{1,...,J\}}(t|\mathbf{M^{(-\mathbf{j})},C})$ to denote the hazards and survival probability, respectively, at time $t$ when exposed to environmental mixture components who’s indices are in $\mathbf{j}$ (referred to as a metal from here on out, although applicable to more general environmental exposures) at their respective $k_1^{th},…,k_{|\mathbf{j}|}^{th}$ percentiles, conditional on all other metals $\mathbf{M^{(-\mathbf{j})}}$ and all confounders $\mathbf{C}$. We quantified the effect of an exposure to an individual metal, $m_j$, on the survival outcome as the HR and the survival probability difference for an interquartile range (IQR) change in exposure to metal $m_j$, holding all other metals and confounders constant. The HR can be thought of as the excess hazard from being exposed to a higher level compared to a lower level of metal $m_j$. The survival probability difference can be thought of as the reduction in survival probability from being exposed to a higher level compared to a lower level of metal $m_j$. We estimated these by taking the ratio of the estimated hazards and the difference of the estimated survival probabilities when $m_j$ was at its $75^{th}$ and $25^{th}$ percentile, while setting the confounders and other metals to their median values. Similarly, we quantified the effect of the metal mixture on the survival outcome as the HR and the survival probability difference for the set of $J$ metals, $m_1, m_2,…,m_J$, all at their $75^{th}$ percentile compared to their $25^{th}$ percentile.

To better understand whether an interaction exists between metals, the interaction was quantified on the multiplicative scale. Define $HR_{j,j'}^{k,k'} = \frac{\lambda_{j,j'}^{k,k'}(t|\mathbf{M}^{\{j,j'\})}, \mathbf{C})}{\lambda_{j,j'}^{k,k}(t|\mathbf{M}^{(-\{j,j'\})}, \mathbf{C})}$. The subscript specifies the two metals we are interested in detecting an interaction between, $m_j$ and $m_{j'}$ where $j \neq j'$. The superscript specifies the percentiles we are interested in detecting an interaction at, this can be thought of as levels of high versus low exposures. The multiplicative interaction can then be defined as $\frac{HR_{j,j'}^{k',k'}}{HR_{j,j'}^{k',k}HR_{j,j'}^{k,k'}}$, or the excess hazard in the presence of high levels of metals $m_j$ and $m_{j'}$ jointly compared to high levels of each metal individually. \cite{vanderweele2014} We estimated these by setting $k=25$ and $k'=75$ and plugging in the estimated hazards into the equation above, while setting the confounders and other metals to their median values.

The chosen estimands for this study are not exhaustive; there are many other estimands that may be of interest to researchers. For example, interaction effects can also be quantified on the additive scale for survival outcome models via the relative excess risk due to interaction ($RERI=HR_{j,j'}^{k',k'}-HR_{j,j'}^{k',k}-HR_{j,j'}^{k,k'}+1$). \cite{VanderWeele2011} However, they wouldn't be properly defined for models that do not allow for interactions, such as Cox PH Model without interactions, because while the product term is forced to be zero between the two metals of interest, the estimated RERI may still be different from zero. On the other hand, the multiplicative interaction will be one when no interaction is included in the model. We therefore chose not to include the RERI in our study.

The mathematical representation of the five estimands we will estimate are shown in Table \ref{tab1}. As seen in the formulations of these estimands, they may vary depending on time, $t$, thus researchers must specify a time, $t_{spec}$, to estimate a point estimate. The choice in $t_{spec}$ should be motivated by contextual relevance. For demonstrative purposes, we considered the $80^{th}$ percentile of the observed total follow-up time, calculated using observations’ time to either event or censoring. 

\begin{table*}[!htb]%
\caption{Estimands of interest.\label{tab1}}%
\begin{tabular}{p{30mm} p{25mm} p{95mm}}
\hline
\textbf{Effect} & \textbf{Scale} & \textbf{Formula} \\
\hline
Individual Metal & Multiplicative & $\frac{\lambda_j^{75} (t|\mathbf{M}^{(-j)},\mathbf{C})}{\lambda_j^{25} (t|\mathbf{M}^{(-j)},\mathbf{C})}$ \\
\\
Individual Metal & Additive & $S_j^{75}(t|\mathbf{M}^{(-j)},\mathbf{C}) - S_j^{25}(t|\mathbf{M}^{(-j)},\mathbf{C})$ \\
\\
Metal Mixture & Multiplicative & $\frac{\lambda_{1,...,J}^{75,...,75}(t|\mathbf{C})}{\lambda_{1,...,J}^{25,...,25}(t|\mathbf{C})}$\\
\\
Metal Mixture & Additive & $S_{1,...,J}^{75,...,75}(t|\mathbf{C}) - S_{1,...,J}^{25,...,25}(t|\mathbf{C})$\\
\\

Interaction & Multiplicative & $\frac{\lambda_{j,j'}^{75,75}(t|\mathbf{M}^{(-\{j,j'\})}, \mathbf{C}) \times \lambda_{j,j'}^{25,25}(t|\mathbf{M}^{(-\{j,j'\})}, \mathbf{C})}{\lambda_{j,j'}^{25,75}(t|\mathbf{M}^{(-\{j,j'\})}, \mathbf{C}) \times \lambda_{j,j'}^{75,25}(t|\mathbf{M}^{(-\{j,j'\})}, \mathbf{C})}$, for $j \neq j'$\\
\hline
\multicolumn{3}{p{155mm}}{\footnotesize $\lambda^{k_1,...,k_{|j|}}_{\mathbf{j}\subset \{1,...,J\}}(t|\mathbf{M^{(-j)},C})$ and $S^{k_1,...,k_{|j|}}_{j\subset \{1,...,J\}}(t|\mathbf{M^{(-\mathbf{j})},C})$ denote the hazards and survival probability, respectively, at time $t$ when exposed to environmental mixture components who’s indices are in $\mathbf{j}$ at their respective $k_1^{th},…,k_{|j|}^{th}$ percentiles, conditional on all other metals $\mathbf{M^{(-j)}}$ and confounders $\mathbf{C}$.}

\end{tabular}
\end{table*}

Environmental epidemiologists are often concerned not only with estimands but with exposure-response functional relationships as well. We further considered a method's ability to capture the true exposure-response relationship between an individual metal and the survival outcome. To do so for survival outcomes, we considered survival probability as a function of the individual metal of interest at the prespecified time $t_{spec}$, holding all other metals and confounders at their median values, $S(m_j|t_{spec},\mathbf{M}^{(-j)},\mathbf{C})$.

\subsection{Statistical Methods}

We conducted a thorough literature search of current machine learning approaches available for survival outcomes. To find machine learning techniques that qualify for our purposes, we searched the key phrases "machine learning for survival analysis”, “high dimensional exposures and survival outcomes”, and “machine learning for environmental mixtures” in Google Scholar. We reviewed the papers and selected a method if it met the following criteria: (1) allows for survival time outcomes; (2) allows for continuous exposures and potential nonlinear effects of these exposures on the outcome; (3) allows for interactions among the exposures; (4) allows for inference (i.e., gives estimates and confidence intervals/standard errors); and (5) is implementable in the statistical computing software R. Additionally, we preferred methods that can perform variable selection and non-parametrically estimate exposure-response relationships by avoiding specification of the functional form of these. To increase the utility of our findings, we chose to additionally include methods that did not necessarily meet all the criteria stated above but have been commonly used in environmental mixtures studies in our comparison.

We found the following methods to meet our criteria: Bayesian Additive Regression Trees (BART), Multivariate Adaptive Regression Splines (MARS), and Gaussian Process Regression (GPR).\cite{Sparapani2016, LeBlanc1999, Williams1998} Additionally, we chose to include the Cox PH model with and without penalized splines and Cox Elastic Net (EN) although they do not necessarily meet our criteria, due to their popularity in environmental mixtures studies. \cite{Cox1972, Malloy2017, Simon2011} Other methods have been extended to the survival outcome setting, but do not meet our criteria or were not applicable for the research questions we were interested in (i.e. do not easily provide survival probability estimates and/or estimates for inference). These included random survival forests, support vector machine, and principal component analysis. \cite{Biganzoli2002, Bair2006, Fouodo2018} Therefore, these were not considered. 

We compared the abilities of the six methods in estimating environmental mixtures effects on a survival outcome. These methods are divided into two categories: those based on the Cox PH model and require the proportional hazards assumption, and those that use a discrete time approach and allow effects to vary with time. Each method is described below; see Table \ref{tab2} for a summary of the characteristics of each method. 

For all non-Bayeisan models, we chose to estimate confidence intervals via the nonparametric bootstrap. While models under the PH approach could rely on the asymptotic distributions of the effect estimates to estimate confidence interva;s, we chose to use the nonparametric bootstrap instead to be more consistent with the other methods we are comparing the Cox PH models to. The bootstrap confidence interval may be more accurate in the presence of model misspecification or a small sample size, both of which are common in environmental mixtures studies. We randomly sampled 100 datasets with replacement of size $n$ (where $n$ is the sample size) from the original dataset and estimated each estimand of interest for each bootstrap sample. The $2.5^{th}$ and $97.5^{th}$ percentiles of these 100 estimated estimands were used as the lower and upper limits, respectively, of the confidence intervals. 

\begin{table*}%
\centering
\caption{Summary of modeling methods.\label{tab2}}%
\begin{tabular}{p{4cm} | p{1.5cm}| p{1.5cm}| p{1.5cm}| p{1.5cm}| p{1.5cm}| p{1.5cm}| p{1.5cm}}
\hline

\textbf{} & \textbf{Time-to-event outcome} & \textbf{Multiple, continuous exposures} & \textbf{Non-linear and interaction effects on the outcome} & \textbf{Inferences (point estimates and confidence intervals)} & \textbf{Variable selection} & \textbf{Automated functional form of covariates} & \textbf{No proportional hazards assumption}  \\

\hline
\textbf{Cox Proportional Hazards Model} & \hfil\checkmark & \hfil\checkmark & & \hfil\checkmark & & & \\
 & & & & & & & \\
\textbf{Cox Proportional Hazards Model with Penalized Splines} & \hfil\checkmark & \hfil\checkmark & \hfil\checkmark & \hfil\checkmark & & & \\
 & & & & & & & \\
\textbf{Cox Proportional Hazards Model with Elastic Net Penalization} & \hfil\checkmark & \hfil\checkmark &  & \hfil\checkmark & \hfil\checkmark & & \\
 & & & & & & & \\
\textbf{Multivariate Adaptive Regression Splines} & \hfil\checkmark & \hfil\checkmark & \hfil\checkmark & \hfil\checkmark & \hfil\checkmark & \hfil\checkmark & \hfil\checkmark\\
 & & & & & & & \\
\textbf{Gaussian Process Regression} & \hfil\checkmark & \hfil\checkmark & \hfil\checkmark & \hfil\checkmark & \hfil\checkmark & \hfil\checkmark& \hfil\checkmark\\
 & & & & & & & \\
\textbf{Bayesian Additive Regression Trees} & \hfil\checkmark & \hfil\checkmark & \hfil\checkmark & \hfil\checkmark & \hfil\checkmark &  \hfil\checkmark & \hfil\checkmark\\

\hline
\end{tabular}
\end{table*}

\subsubsection{Approach 1 - Proportional hazards models}

Cox PH and its extensions are commonly used for survival outcomes. The simultaneous effect of several variables on the survival outcome at a specified time, $t$, is estimated by assuming the hazard function has the form 

$$\lambda(t; \mathbf{m,c}) = \lambda_0 (t) exp \{f(\mathbf{m,c})\}$$

\noindent where $\lambda_0(t)$ is the baseline hazard function, or the hazard at time $t$ when all metals and continuous confounders are equal to 0 and all categorical confounders are equal to their reference group. While $\lambda_0 (t)$ can vary over time, the hazard of the event given a set, $\{\mathbf{m}_i,\mathbf{c}_i\}$ is a constant multiple of the hazard for another set, $\{\mathbf{m}_{i^*},\mathbf{c}_{i^*}\}$ where $i\neq i^*$. This is referred to as the proportional hazards assumption. This assumption implies that the ratio of the hazards for any two exposure profiles is constant over time, therefore the HR is independent of $t$. \\

\noindent\textbf{Traditional Cox Proportional Hazards Model (Cox PH)}

\noindent For the Cox PH model, the hazard function is assumed to have the form

$$\lambda(t;\mathbf{m,c}) = \lambda_0 (t) exp\{\Sigma+{j=1}^J \beta_j m_j + \Sigma_{l=1}^L \gamma_l c_l\},$$

\noindent for $J$ metals and $L$ confounders. \cite{Cox1972} The coefficients are estimated via the partial likelihood method, which was performed using the R package \textit{survival} (version 3.2-7). To increase comparability with the other methods, a model with two-way interactions between all metals was also included. Although among the most common models used, this model does not perform automated variable selection or employ any procedure to mitigate the impact of multicollinearity of multiple predictors common in mixture analyses. Thus, in scenarios where the mixture components are highly correlated and the number of components in the mixture set is large relative to the sample size, we may see suboptimal performance. Additionally, it assumes a linear relationship with the outcome of interest. \\

\noindent\textbf{Cox Proportional Hazards Model with Penalized Splines (Cox PH-ps)}

\noindent We considered an extension of the traditional Cox PH model and include penalized splines to allow for nonlinear effects.\cite{Wood2016} The hazard function has the form

$$\lambda(t;\mathbf{m,c}) = \lambda_0 (t) exp\{\Sigma_{j=1}^J f_j(m_j) + \Sigma_{j \neq j'}f_{j, j'}(m_j, m_{j'}) + \Sigma_{l=1}^L \gamma_l c_l \},$$

\noindent where $f_j$ and $f_{j, j'}$ are smooth functions estimated via penalized splines through tensor product smoothers, estimated using the R package \textit{mgcv} (version 1.8-33). The same assumptions hold as with the traditional model, however the estimated effect of each metal and their interaction with other metals on the survival outcome is more flexible. A downside to this approach is it requires one to specify the functional form (i.e. specify which metals/interactions should be entered nonlinearly) and tends to be less computationally and statistically efficient due to the fact it is less parsimonious than the traditional Cox PH model. We considered all metals and two-way interactions between metals ($\mathbf{m}$) to be more flexible while all confounders ($\mathbf{c}$) were entered linearly. \\

\noindent\textbf{Cox Elastic Net (Cox EN)}

\noindent EN is a popular approach for high-dimensional, correlated data. For the survival analysis extension, the form of the estimated hazard function is the same as that of the traditional Cox PH model. However, the estimation method maximizes the likelihood subject to the constraint $\omega \Sigma |\beta_j| + (1 - \omega) \Sigma \beta_j^2 \leq d$. \cite{Simon2011} $\omega \in [0,1]$ is the mixing parameter which specifies the percentage of the constraint pertaining to the L1 ($\Sigma |\beta_j|$) and L2 ($\Sigma \beta_j^2$) penalties. To estimate $\beta$, the optimization problem subject to the given constraint can be written as its Lagrangian formulation and we consider the regularization parameter $\kappa \in [0,\infty)$. If $\kappa = 0$ then this is equivalent to the traditional Cox PH model. The larger the $\kappa$, the stronger the regularization and the more the coefficients will be shrunk towards zero. In our model, the coefficients pertaining to the confounders ($\gamma$) were not included in the constraint. As with the traditional Cox PH model, Cox EN with two-way interactions between metals was also considered.

Cross-validation (CV) was used to find the optimal choices for $\omega$ and $\kappa$. The R package \textit{glmnet} (version 4.1-3) incorporates k-fold cross-validation when estimating the Cox EN coefficients. The cv.glmnet command returns a value of $\kappa$ for a fixed $\omega$. A pathwise solution via cyclic coordinate descent is used, where $\kappa$ ranges from sufficiently large enough for $\beta = 0$, and decreases until near the unregularized solution ($\kappa = 0$). To find the optimal $\alpha$, a grid of potential values was used (see Table \ref{tab3}). At each value of $\omega$, the optimal $\kappa$ is determined through the CV described. The final model is the combination of $\omega$ and $\kappa$ which maximizes $l(\beta_{-i} (\omega,\kappa))- l_{-i} (\beta_{-i} (\omega,\kappa))$, where $l_{-i}$ is the log partial likelihood excluding part $i$ of the data and $\beta_{-i} (\omega,\kappa)$ is the optimal $\beta$ for the non-left out data. CV was performed for each bootstrap sample when estimating the corresponding confidence interval.

\subsubsection{Approach 2 – Discrete-time survival analysis models}

Many machine learning approaches are yet to be extended to survival data and/or are not readily available in software commonly used by environmental epidemiologists. In order to employ such methods and relax the proportional hazards assumption, we applied a discrete-time survival analysis approach.\cite{Fahrmeir2005} Each subject has a corresponding triple $(t_i,\delta_i,\mathbf{x}_i)$. For subject $i$, $i=1,…,n$, $t_i$ denotes the observed time to event or censoring, $\delta_i$ denotes the indicator of whether the event occurred or was censored, and $\mathbf{x}_i$ denotes the set of observed metals and confounders.

Time, $t$, is discretized into $R$ bins of event/censoring times delineated by its corresponding $1/R,2/R,…,R/R$ quantiles. An augmented dataset is then created such that subject $i$ has multiple corresponding observations for each distinct discretized time up to the bin pertaining to their observed $t_i$. An observation now refers to a subject-time, where confounders/metals are repeated for the number of rows corresponding to each subject in the new augmented dataset. A binary variable, $\mathbf{Y}_i$, is created to indicate event status per subject-discretized time period. For example, if $R=3$ and time is discretized into [0,1), [1,2), and [2,3) bins then if subject $i$ experienced the event at time 1.5, they will have two observations in the augmented dataset where $\mathbf{Y}_i=(0,1)$. Conversely, had this subject been censored at time 1.5 then there would still be two observations in the augmented dataset, however $\mathbf{Y}_i = (0,0)$. For more detail and an example see (Sparapani et al, 2016).\cite{Sparapani2016}

$\mathbf{Y}$ is now the outcome of interest and one can apply methods for binary outcomes to the augmented dataset with $t$ included as a confounder. Generally, we model

$$g(\mu_i) = f(\mathbf{m_i,c_i},t),$$

\noindent where $g(.)$ is a monotonic link function, $\mu_i=E(Y|\mathbf{m_i,c_i},t)$ is the expected probability of experiencing the event, and $f(\mathbf{m_i,c_i},t)$ is a function of the metals, confounders and time estimated through an algorithm. We selected algorithms which flexibly fit nonlinear response surfaces even with a large number of predictors without requiring the researcher to specify variable importance or the functional form of the relationship between predictors and the outcome. We used $R=5$ for all our discrete time models.

To estimate the estimands of interest using the discrete time approach for survival outcomes, one can estimate the survival probability at discretized time bin $r$, for $r=1,…,R$, as $S(t_{(r)}) | \mathbf{m,c}) = Pr(T > t_{(r)} | \mathbf{m,c}) = \Pi_{l=1}^r (1 - pr(t_{(l)}, \mathbf{m,c}))$ and the hazard as $\lambda(t_{(r)} | \mathbf{m,c}) = \frac{pr(t_{(r)},\mathbf{m,c})}{(t_{(r)}-t_{(r-1)})}$.  Here, $pr(t_{(r)},\mathbf{m,c})$ denotes the probability of experiencing the event at time $t_{(r)}$, with exposure fixed to $\mathbf{m}$ and confounders $\mathbf{c}$ conditional on no previous event. \\

\noindent\textbf{Multivariate Adaptive Regression Splines (MARS)}

\noindent The MARS algorithm creates a piecewise linear model.\cite{Friedman1995} Nonlinear relationships can be captured in the data by binning the range of values for each exposure into smaller sections, split by values referred to as knots, and creating linear regression models for each section. CV is performed to optimize the maximal degree of interactions ($D$) and the number of terms retained ($P$) in the final model, which has the form

$$f(\mathbf{m,c},t)=\Sigma_{p=1}^{P} \beta_p h_p (\mathbf{m},t) + \Sigma_{l=1}^L \gamma_l c_l,$$

\noindent The confounders are included in the model linearly, as seen in $\Sigma_{l=1}^L \gamma_l c_{l}$. $h_p (\mathbf{m},t)$ is a piecewise linear function of a metal (or interaction of piecewise linear functions) included in the final model. For example, $h_p (\mathbf{m},t)=(m_j-\circ)_+$ is a piecewise linear function of metal $m_j$ starting from the knot value $\circ$. The “+” indicates that we are referring to the positive section, such that $h_p (\mathbf{m},t) = m_j - \circ$ when $m_j>\circ$ and 0 otherwise. $h_p (\mathbf{m},t)$ can also be a product of piecewise linear functions, such as $h_p (\mathbf{m},t) = (m_j - \circ_1 )_+ \times (m_{j^{'}} - \circ_2)_+$. We used the logit link function for our MARS model adaptation to discrete time survival analysis.  MARS is a convenient approach for modeling nonlinear and interaction relationships of the metals while still maintaining some interpretability. However, it can be computationally difficult, and a lot of power would be needed to consider a model with more than two degrees of interaction ($D>2$).

We used the R package \textit{earth} (version 5.3.1) and perform 5-fold CV for $D$ (using grid of values: [1,2]) and $P$ (using grid of values: [5, 10, …, 50]) using the R package \textit{caret} (version 6.0-90). The area under the receiver operating characteristic (ROC) curve is used for goodness of fit in the CV procedure. CV was performed for each bootstrap sample when estimating the corresponding confidence interval. \\

\noindent\textbf{Gaussian Process Regression (GPR)}

\noindent GPR is a nonparametric method which flexibly models the combined effects of metals and confounders. Rather than imposing a functional form on $f(\mathbf{m,c},t)$, GPR estimates the outcome of interest for a given set of values $\mathbf{\chi}_i = \{\mathbf{m_i,c_i},t\}$ by weighting all observations based on how numerically close the confounders/metals are to these specified values. To compute these weights, GPR uses a Gaussian kernel to measure the distance between $\mathbf{\chi}_i$ and $\mathbf{\chi}_{i'}$ for $i \neq i'$. The distance is measured as $K_{ii^{'}}(\chi_i, \chi_{i^{'}}) = exp\{-\frac{1}{\rho} \Sigma_{l=1}^{|\mathbf{\chi}|}  (\chi_{il} - \chi_{i^{'}l})^2 \}$, where $|\mathbf{\chi}|=J+L+1$, the number of components in $\mathbf{\chi}$. Each subject from the training dataset receives a weight, such as $w_{ii'} = \frac{K_{ii'}}{\Sigma_{i'=1}^n K_{ii'}}$, which increases as the distance decreases. Observations with similar exposure profiles should have a similar outcome, thus their $K_{ii'}$ should be smaller and thus receive higher weights. The estimated outcome can be thought of as a weighted average of the observed outcomes, $g(\hat{\mu}_i) = \Sigma_{i'=1}^n w_{ii'} y_{i'}$. We used the logit link function to apply the discrete time survival analysis framework. Since a functional form is not specified or assumed, nonlinear exposure-response functions and nonlinear and non-additive interactions among all mixture components can be captured.

$\rho$ is a tuning parameter used to control flexibility, as seen in the smoothness of the exposure-response curve. Given the time intensiveness to run GPR, we used the median of $|\chi - \chi'|^2$ for the $\rho$ value rather than cross-validating. While this might not provide the optimal $\rho$, it has been shown empirically that the optimal value lies between the 0.1 and 0.9 quantiles of this statistic. \cite{Caputo2001} To fit the model, we used the R package \textit{kernlab} (version 0.9-29).\\

\noindent\textbf{Bayesian Additive Regression Tree (BART)}

\noindent BART is a nonparametric Bayesian regression method which approximates $f(\mathbf{m_i,c_i},t)$ using a sum of trees approach.\cite{Chipman2010} $h_p (\mathbf{m_i,c_i},t;U_p,V_p)$ represents an individual tree where $p = 1,…,P$. $U_p$ denotes the $p^{th}$  tree’s terminal nodes and $V_p$ denotes its predicted outcome value from the terminal nodes. The function form is

$$f(\mathbf{m_i,c_i},t) = \Sigma_{p=1}^P h_p (\mathbf{m_i,c_i},t;U_p,V_p).$$

\noindent A Bayesian approach is used to fit the trees. A prior is imposed to regularize the fit by keeping individual tree effects small. Interactions and nonlinearities are naturally incorporated into the tree structure. By imposing a Bayesian framework, credible intervals are also easily produced.

To use BART for the survival setting, we used the probit link function.\cite{Sparapani2016} The prior follows $pr((U_1,V_1  ),…,(U_P,V_P))= \Pi_p pr(U_p)[\Pi_i  pr(\mu_{ip}  |U_p)]$. $pr(U_j)$ is specified by (i) the probability that a depth $d$ node is nonterminal, (ii) the distribution on the splitting variable assignments at each interior node, and (iii) the distribution on the splitting rule assignment in each interior node, conditional on the splitting variable. (i) is given by $a(1+d)^{-b}$, $a \in (0,1)$ and $b \in [0,\infty)$. The software default values for the hyperparameters are $a = 0.95$ and $b = 2$. Using these values, trees with $1, 2, 3, 4$ and $\geq 5$ terminal nodes receive prior probability of 0.05, 0.55, 0.28, 0.09 and 0.03, respectively, such that individual trees are kept small. A uniform prior is placed on the choice of splitting variable at each node, and a discrete uniform prior is specified for the splitting values. For $p(\mu_{ij}|U_j)$ we consider $\mu_{ij} \sim N(0,\sigma_\mu^2)$ where $\sigma_\mu = 3.0/(k\sqrt{P})$. This prior shrinks the tree parameters $\mu_{ip}$ toward zero, limiting the effect of individual tree components. As $k$ and/or the number of trees $P$ increases, this prior will become tighter and apply greater shrinkage to $\mu_{ip}$.

We fit BART using the R package \textit{BART} (version 2.9.0). The default parameter values provided by the package, $a = 0.95$, $b = 2$, $k = 2$ and $P=50$, have been shown to perform well. Thus, we choose to use them in all our analyses to aid run time at the expense of potential improvements to model performance. Alternatively, the value of $k$ and $P$ may be chosen by CV. To sample from the posterior distribution, a Bayesian back-fitting MCMC algorithm was used. 1000 draws from the posterior were generated after a burn-in of 250 draws. Due to the repetition between observations of the augmented dataset, we are more likely to run into issues with high auto-correlation, thus we used thinning of 250 between each returned value.

\begin{table}%
\centering
\caption{Methods' choice of values for required hyperparameters and run time, including bootstrapping.\label{tab3}}%
\begin{tabular}{p{5cm} | p{1.5cm} | p{1.5cm} | p{3cm} | p{3cm}}
\hline
\multicolumn{2}{c|}{\textbf{Method}} & \textbf{Hyper-parameter} & \textbf{Grid of Values Considered} & \textbf{Average Run Time (IQR), minutes} \\
\hline
\multirow{2}{*}{\textbf{Cox Proportional Hazards Model}} & \multirow{2}{*}{Cox PH} & \multirow{2}{*}{none} &  \multirow{2}{*}{ }  & 0.99 (0.92, 1.10) \\
& & & & 1.48 (1.42, 1.52)* \\
\hline
\textbf{Cox Proportional Hazards Model with Penalized Splines} & Cox PH-ps & none &  & 1.56 (1.18, 1.43)\\
\hline
\multirow{2}{5cm}{\textbf{Cox Proportional Hazards Model with Elastic Net Penalization}} & \multirow{2}{*}{Cox EN} & $\omega$ &  $\{0.0, 0.2, ..., 1.0\}$  & 3.68 (3.12, 4.31) \\
\cline{3-4}
& & $\kappa$ & Pathwise solution& 19.9 (12.8, 30.8)* \\
\hline
\multirow{2}{5cm}{\textbf{Multivariate adaptive regression splines}} & \multirow{2}{*}{MARS} & $P$  & \{5, 10, ..., 50\} & 50.6 (51.7, 55.7)\\
\cline{3-4}
& & $D$ & \{1,2\} &  \\
\hline
\textbf{Gaussian process regression} & GPR & $\rho$ & Data driven & 237.0 (219.0, 249)\\
\hline
\multirow{4}{*}{\textbf{Bayesian additive regression trees}} & \multirow{4}{*}{BART} & $P$  & 50 & 30.2 (29.3, 31.4)\\
\cline{3-4}
& & $k$ & 2 &  \\
\cline{3-4}
& & $a$ & 0.95 &  \\
\cline{3-4}
& & $b$ & 2 &  \\
\hline
\multicolumn{5}{{p{\linewidth}}}{\footnotesize *Average run time for the model with interaction terms.}
\end{tabular}
\end{table}

\subsection{Simulation Study}

We conducted simulations mimicking the data observed from the SHS cohort to quantify the performance of each method under different real-world scenarios. We considered various scenarios: (1) metals have a linear effect  on time-to-event outcome on the log-hazard scale and the proportional hazards assumption holds (i.e. the Cox PH model is correctly specified), mixture consists of 5 components, (2) metals have a nonlinear effect on time-to-event outcome and the PH assumption holds, mixture consists of 5 components, (3) metals have a linear effect on time-to-event outcome and the proportional hazards assumption is violated, mixture consists of 5 components, and (4) same as scenario 1 but with 10 mixture components. The data were generated such that the correlations across mixture components are similar to the SHS, which are low to moderate. Thus we also included a scenario (5) where everything is equivalent to scenario 1, but with higher correlations across mixture components. 

Table \ref{tab4} lists the different simulation scenarios and specifies our choices in parameter specifications, explained in more detail below. Although not exhaustive, these scenarios were chosen to simulate common real-world scenarios. For each scenario, we simulated $F = 400$ datasets each with $n=1000$ observations to replicate a common number of observations in prospective epidemiological studies. The supplementary material includes simulation results considering selected scenarios with $n=3000$ observations to closer replicate the SHS.

\begin{table*}%
\caption{Simulation study scenarios.\label{tab4}}%
\begin{tabular}{p{1cm} | p{3cm} | p{4cm} | p{4cm} | p{2cm}}
\hline
\textbf{\#} & \textbf{Scenario} & $\alpha(\mathbf{m})$ & $g(\mathbf{m})$ & \textbf{\# of components in metal mixture (J)} \\
\hline
1 & Base Case & 1 & $exp\{\Sigma_{j=1}^J \beta_j m_j\}$ & 5 \\
 & & & & \\

2 & Nonlinear & 1 & $-1.55(m_1 + 2)^{1/4} + \frac{8}{1 + exp\{3.3m_3 -7\}} + 1.5(m_4 + 3.5^2(m_5 + 1))$  & 5 \\
 & & & & \\

3 & Proportional hazards violation & $0.7+0.1m_1+0.1m_3+0.1m_4+0.1m_5$ & $exp\{\Sigma_{j=1}^J \beta_j m_j\}$ & 5 \\
 & & & & \\

4 & Higher correlations across mixture components & 1 & $exp\{\Sigma_{j=1}^J \beta_j m_j\}$ & 5 \\
 & & & & \\

5 & Higher dimensionality of mixture & 1 & $exp\{\Sigma_{j=1}^J \beta_j m_j\}$ & 10 \\
\hline
\end{tabular}
\end{table*}

\subsubsection{Simulating exposure data}

We simulated confounders and environmental mixture data such that the distributions and correlation structure closely mimic that of the SHS. For all scenarios, three confounders were simulated: $sex \sim Bernoulli(p=0.59)$,  $BMI\sim N(3.39,0.19)$, and $age \sim N(56.13,8.10)$. Let $\mathbf{C} =\{Sex,BMI,Age\}$ denote the set of simulated confounders.

The metals were simulated as a linear function of these confounders, previously simulated metals, and noise. For example:

$$M_1 = \Phi_{0,1}+ \Phi_{sex,1} sex + \Phi_{BMI,1} BMI + \Phi_{age,1} age+N(0,\sigma_1^2)$$
$$M_2=\Phi_{0,2}+\Phi_{sex,2} sex+\Phi_{BMI,2} BMI+\Phi_{age,2} age+\Phi_{1,2} M_1+N(0,\sigma_2^2)$$
$$...$$
$$M_J=\Phi_{0,J}+\Phi_{sex,J} sex+\Phi_{BMI,J} BMI+\Phi_{age,J} age+\Phi_{1,J} M_1+\Phi_{2,J}M_2+ ...+ \Phi_{J-1, J} M_{J-1}+N(0,\sigma_J^2)$$

\noindent $J$ is the number of metals simulated for the exposure mixture set $M=\{M_1,M_2,...,M_J\}$. The coefficients used for the linear functions came from the coefficients estimated from linear regression models for the corresponding variables in the SHS data (based on log-transformed metals) and their respective $\sigma_j^2$. This approach leads to correlations similar to that observed in the SHS, where correlations across metals range from 0.00 to 0.26. For the scenario with higher correlations, the coefficients pertaining to the metals were higher, resulting in correlations closer to 0.40. The correlation matrices for the SHS cohort data and examples from the simulated datasets are provided in the supplementary material. 

\subsubsection{Simulating survival outcome}

The time-to-event outcome was simulated as $t \sim Weibull(\alpha(\mathbf{m,c}),f(\mathbf{m,c}))$, where 

$$S(t;\mathbf{m,c})= exp\{-(\frac{t}{f(\mathbf{m,c})} )^{\alpha(\mathbf{m,c})}\}$$
$$\lambda(t;\mathbf{m,c}) = \frac{\alpha(\mathbf{m,c})}{f(\mathbf{m,c})}(\frac{t}{f(\mathbf{m,c})})^{(\alpha(\mathbf{m,c})-1)}$$

To generate a distribution of survival times similar to that observed in the SHS, two censoring random variables were used, $C_1 \sim Uniform(0,100)$ and $C_2 \sim Uniform(16,20)$. This setting replicates the scenario observed in the SHS in which there is continuous uniform censoring followed by increased uniform censoring towards the end of follow-up time. All scenarios approximate the amount of censoring observed in the SHS, where 67\% of observations do not experience the event by the end of their follow-up time. We also investigated the performance of each method under no censoring, the results for which can be seen in the supplementary material.

We controlled whether the proportional hazards assumption holds through $\alpha(\mathbf{m,c})$ and the functional form of the effect of $\mathbf{C}$ and $\mathbf{M}$ on the outcome through $f(\mathbf{m,c})$. For all scenarios, the confounders $\mathbf{C}$ were considered to have a linear effect on the outcome, such that $f(\mathbf{m,c}) = exp\{\Sigma_{k=1}^3 \gamma_k c_k + g(\mathbf{m})\}$, and no effect on $\alpha$, such that $\alpha(\mathbf{m,c})=\alpha(\mathbf{m})$. Only four components of $\mathbf{M}$ were included in $f(\mathbf{m,c})$ or $\alpha(\mathbf{m})$ such that not all metals were associated with the outcome. All functional forms and parameter values were chosen such that survival time outcomes and the percentage of observations that were censored were similar to that observed in the SHS. The specifications for $g(\mathbf{m})$ and $\alpha(\mathbf{m})$ for each scenario are shown in Table \ref{tab4}.

\subsubsection{Estimating methods' accuracy}

For the simulation study, methods were compared in terms of their bias and efficiency in estimating the true values of the estimands described (individual, mixture, and interaction effects). We selected $t_{spec}$ to be the $80^{th}$ percentile of the time to event/censoring, such that we quantified how accurate the methods were at estimating the estimands at the $80^{th}$ percentile of $t$. The relative bias for each method was defined as $\frac{1}{F} \Sigma_{f=1}^{F} (\Omega_{truth}-\hat{\Omega}_f)/\Omega_{truth}$, where $\Omega_{truth}$  refers to the true value of the quantity of interest and $\hat{\Omega}_f$ refers to the estimated value for the quantity of interest using the $f=1,…,F$ simulated dataset. The standard deviation (SD) for a given simulated estimate was estimated either by taking the SD of the estimates from the bootstrap samples or, in the case of the Bayesian methods, the SD of the posterior distribution of the model parameter of interest. For each method, the corresponding root mean squared error was computed as $RMSE = \sqrt{}\frac{1}{F} \Sigma_{f=1}^{F} (\Omega_{truth}-\hat{\Omega}_f)^2$. Coverage was computed as the percentage of the confidence intervals estimated for the specified estimand that included the true value, $\frac{1}{F} \Sigma_{f=1}^F I(\hat{\Omega}_{f,ll} \leq \Omega_{truth} \leq \hat{\Omega}_{f,ul})$ where $\hat{\Omega}_{f,ll}$ and $\hat{\Omega}_{f,ul}$ refer to the estimated lower and upper limits of the confidence/credible intervals using the $f$ simulated dataset.

In addition to the ability of a method to estimate the estimands of interest, we were also interested in estimation of the exposure-response curve. We considered the survival probability over a range of exposure values for an individual metal, holding all other metals/confounders at their median. To quantify the ability of a method to estimate this curve, we used the mean integrated squared error, which we quantified as $MISE = \frac{1}{F} \Sigma_{f=1}^{F}  \frac{1}{19} \Sigma_{k= 0.05,0.10,...,0.95} (\hat{S}_{f,j}^k (t | \mathbf{M}^{(-j) },\mathbf{C})- S_j^k (t|\mathbf{M}^{(-j)},\mathbf{C}))^2$ where 19 is the length of \{0.05,0.10,…,0.95\}. $\hat{S}_{f,j}^k (t | \mathbf{M}^{(-j) },\mathbf{C})$ refers to the estimated survival probability at time $t$ using the $f^{th}$ simulated dataset when exposed to environmental mixture component $j$ at it’s $k^{th}$ percentiles, conditional on all other metals $M^{(-j)}$ and all confounders $C$.

\subsection{Analysis of Strong Heart Study Data}

We used data from the Strong Heart Study (SHS) to demonstrate the applicability of the methods on a real-world example of multiple exposures effect on a survival outcome. Previous assessments of the potential role of metals on CVD have found that urinary arsenic and cadmium levels are prospectively associated with incident clinical CVD in the SHS population.\cite{Moon2013, Tellez2013} A possible interaction between tungsten and molybdenum was also found, with an increased risk of incident CVD associated with tungsten only at low urinary levels of molybdenum.\cite{Nigra2018} Selenium has also been found to be associated with incident CVD, where the exposure-response is nonlinear with both lower and higher levels of urinary selenium associated with excess CVD risk.\cite{Zhao2022} We leveraged the fact that all variables needed for this study, including metals, clinical CVD outcomes, and other relevant variables for model adjustment are available and have been well characterized in multiple publications, such as those aforementioned, to better demonstrate the applicability of the methods beyond simulation studies.

\subsubsection{Study population}

The SHS population has been described in detail previously.\cite{Lee1990} Between 1989 and 1991, 4,549 men and women aged 45–75 from 13 tribes located in Arizona, Oklahoma, North Dakota, and South Dakota were recruited to participate in a prospective cohort study to investigate CVD and its risk factors in American Indian adults. Participants were followed for clinical events through 2017. One community withdrew their consent to participate in further research in 2016, leaving 3,516 participants. Additionally, we excluded 251 individuals with CVD at baseline, 434 without urinary metal measurements, and 103 who were missing other variables of interest, leaving an analytical sample of 2,728 individuals. The study protocol was approved by the institutional review boards of the Indian Health Service, the participating institutions, and the participating tribes. Each participant provided individual written informed consent.

\subsubsection{Metals mixture exposure}

Baseline spot urine samples were collected; detailed analytical methods can be found in Scheer et al., 2012.\cite{Scheer2012} For arsenic, species were first separated using high performance liquid chromatography coupled with inductively coupled plasma mass spectrometry. We used the sum of inorganic and methylated arsenic species as the biomarker of arsenic exposure. The limits of detection (LOD) for arsenic [As], cadmium [Cd], molybdenum [Mo], selenium [Se], tungsten [W], and zinc [Zn] were 0.1, 0.015, 0.1, 2.0, 0.005, and 10.0 µg/L, respectively. The percentage of samples below the LOD in the analytical sample was 0\%, 0.037\%, 0\%, 0\%, 1.2\%, and 0\% for As, Cd, Mo, Se, W, and Zn, respectively. Urinary metal/metalloid concentrations below the LOD were replaced with the LOD divided by the square root of two. To account for urine dilution, urinary metal/metalloid concentrations were divided by urine creatinine levels and expresses as µg/g creatinine. Urinary levels of the metals/metalloids were skewed, thus were log-transformed for all statistical analysis. 

\subsubsection{Cardiovascular disease outcome}

Cardiovascular endpoints were assessed by questionnaire, tribal records, Indian Health Service hospital records, death certificates, and direct contact. All deaths and cardiovascular outcomes were reviewed by the Morbidity and Mortality Review Committee based on the World Health Organization (WHO) criteria. Follow-up through 2017 was 99.8\% complete for mortality and 99.2\% complete for nonfatal events. Incident CVD was defined as definite or possible fatal or non-fatal congenital heart disease (CHD), stroke, or heart failure.\cite{Lee1990, Gillum1984} Survival time was defined as the time from start of follow-up (1989-1991) and was calculated as the difference between age at the date of the baseline examination and age at the date of the cardiovascular event, age at the date of death, or age at end of follow-up, whichever occurred first. The mean follow-up time among participants who did not develop a cardiovascular event over follow-up was 15 years.

\subsubsection{Confounders}

Trained and certified nurses and medical examiners collected information on baseline covariates. To adjust for confounding, the following confounders were included in all models: baseline age, sex, education level (none, some, or high school), smoking status (never, former, or current), body mass index (BMI) (kg/m$^2$), kidney function, and arsenobetaine. Kidney function was quantified as the estimated glomerular filtration rate (eGFR) calculated from creatinine, age, and sex using the Chronic Kidney Disease Epidemiology Collaboration formula.\cite{levey2009} Arsenobetaine was included as a marker of seafood arsenicals. Confounders were entered linearly into all models where the confounders/outcome relationship could be specified.

\section{Results}

\subsection{Simulation Study}

\subsubsection{Estimation accuracy}

We summarise simulation results for scenarios 1-3, a summary of results for scenarios 4 and 5 can be found in the supplementary material. For the simulated metals, the ($25^{th},75^{th}$) percentiles of $M_1,…,M_5$ used for estimating the estimands of interest were (1.60, 2.74), (-0.66, 0.59), (2.78, 3.92), (3.39, 4.46) and (-2.96, -1.20), respectively. The three simulation scenarios simulated the metals equivalently, thus these were the same across the scenarios. The $80^{th}$ percentile of $t$ used as $t_{spec}$ was 18.5, 18.5 and 18.4 years, respectively, for each of the first three scenarios.

\begin{figure*}
\centerline{\includegraphics[scale=0.05]{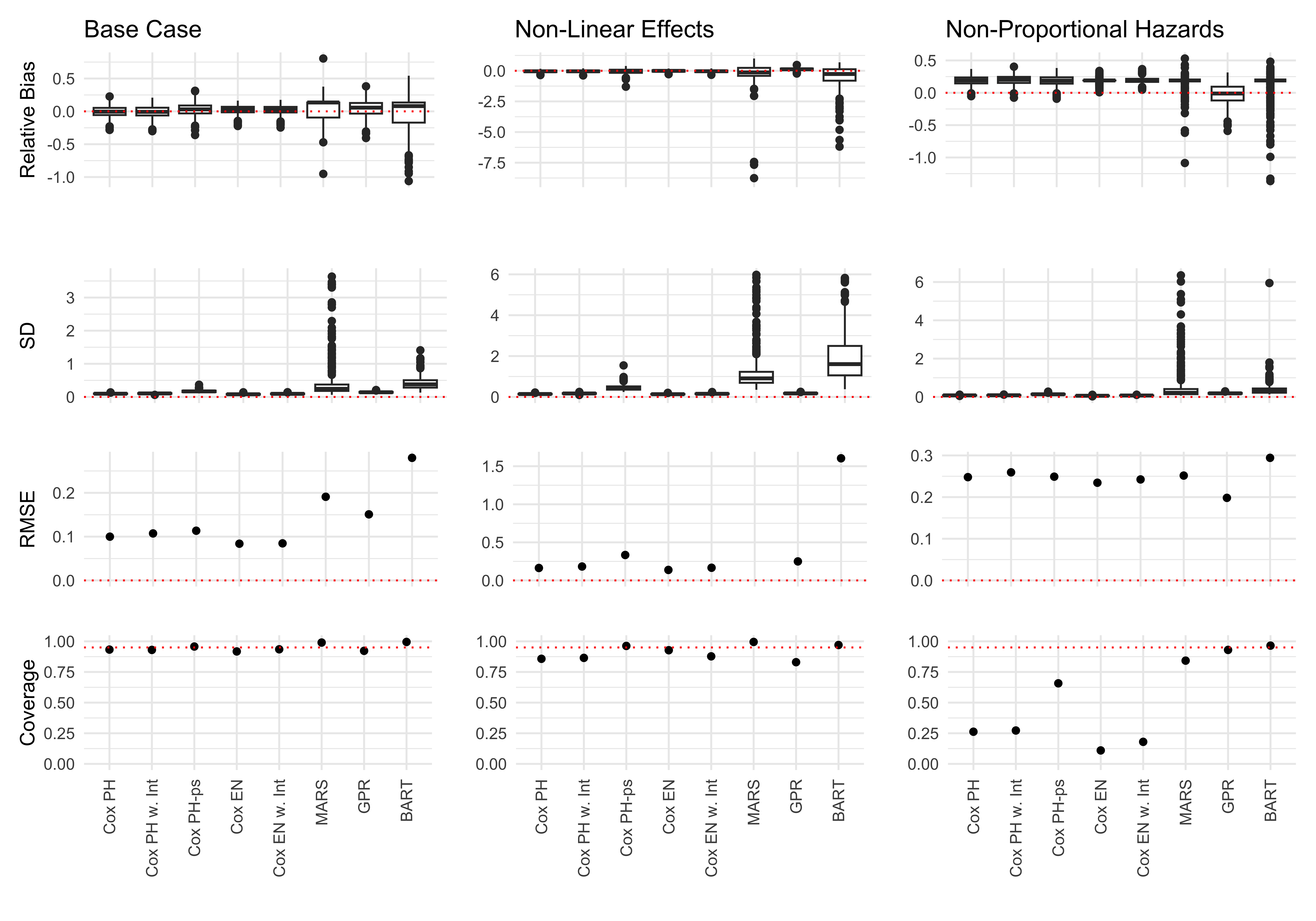}}
\caption{Methods’ performances in estimating an individual metal’s effect on the hazard ratio scale. Extreme values were seen for MARS and BART and are cut off. For Scenario 1, this includes 27 observations of SD for MARS. For Scenario 2, this includes 2 observations of relative bias for MARS, 92 observations for SD for MARS and the RMSE of MARS as well as 21 observations of SD for BART. For scenario 3, this includes 10 observations of SD for MARS.
\label{fig1}}
\end{figure*}

\begin{figure*}
\includegraphics[scale=0.05]{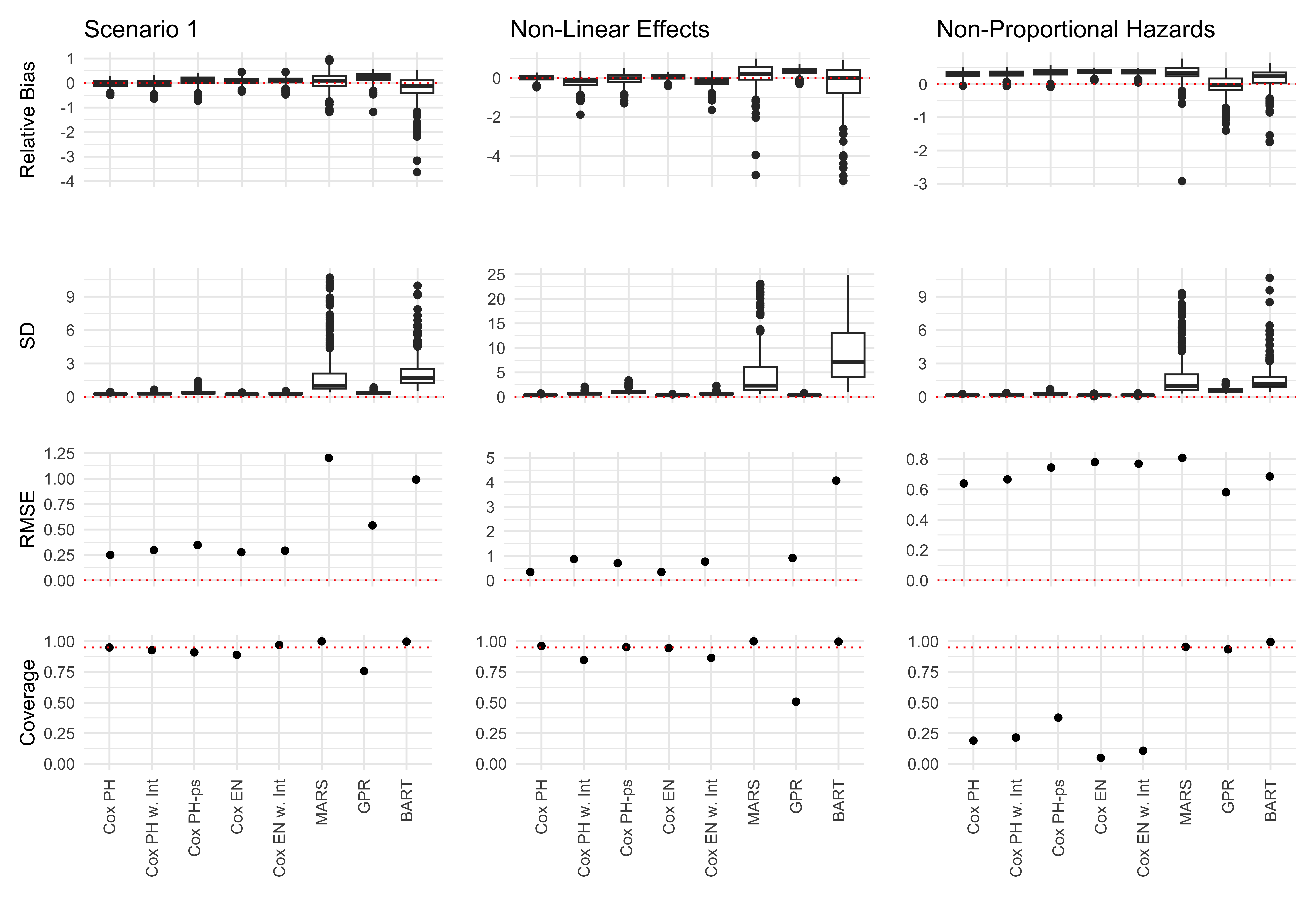}
\centering
\caption{Methods’ performances in estimating an environmental mixture’s effect on the hazard ratio scale. Extreme values were seen for MARS and BART and are cut off. For Scenario 1, this includes 1 observation of relative bias and 27 observations of SD for MARS. For Scenario 2, this includes 6 observations of relative bias, 205 for SD and the RMSE (100,781) of MARS as well as 3 observations of relative bias and 62 of SD for BART. For scenario 3, this includes 19 observations of SD for MARS.
\label{fig2}}
\end{figure*}

\begin{figure*}
\includegraphics[scale=0.05]{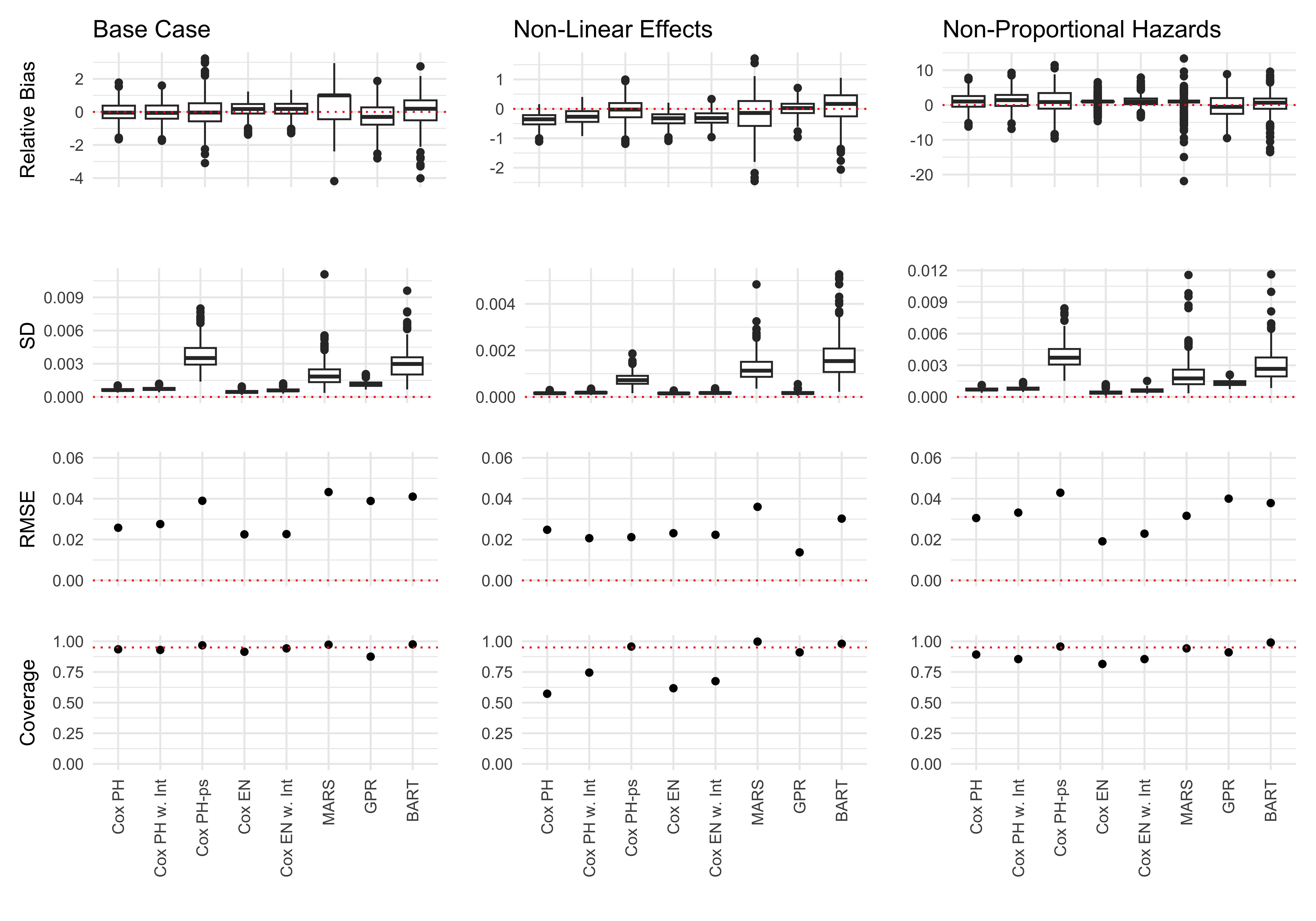}
\centering
\caption{Methods’ performances in estimating an individual metal’s effect on the survival probability difference scale.\label{fig3}}
\end{figure*}

\begin{figure*}
\includegraphics[scale=0.05]{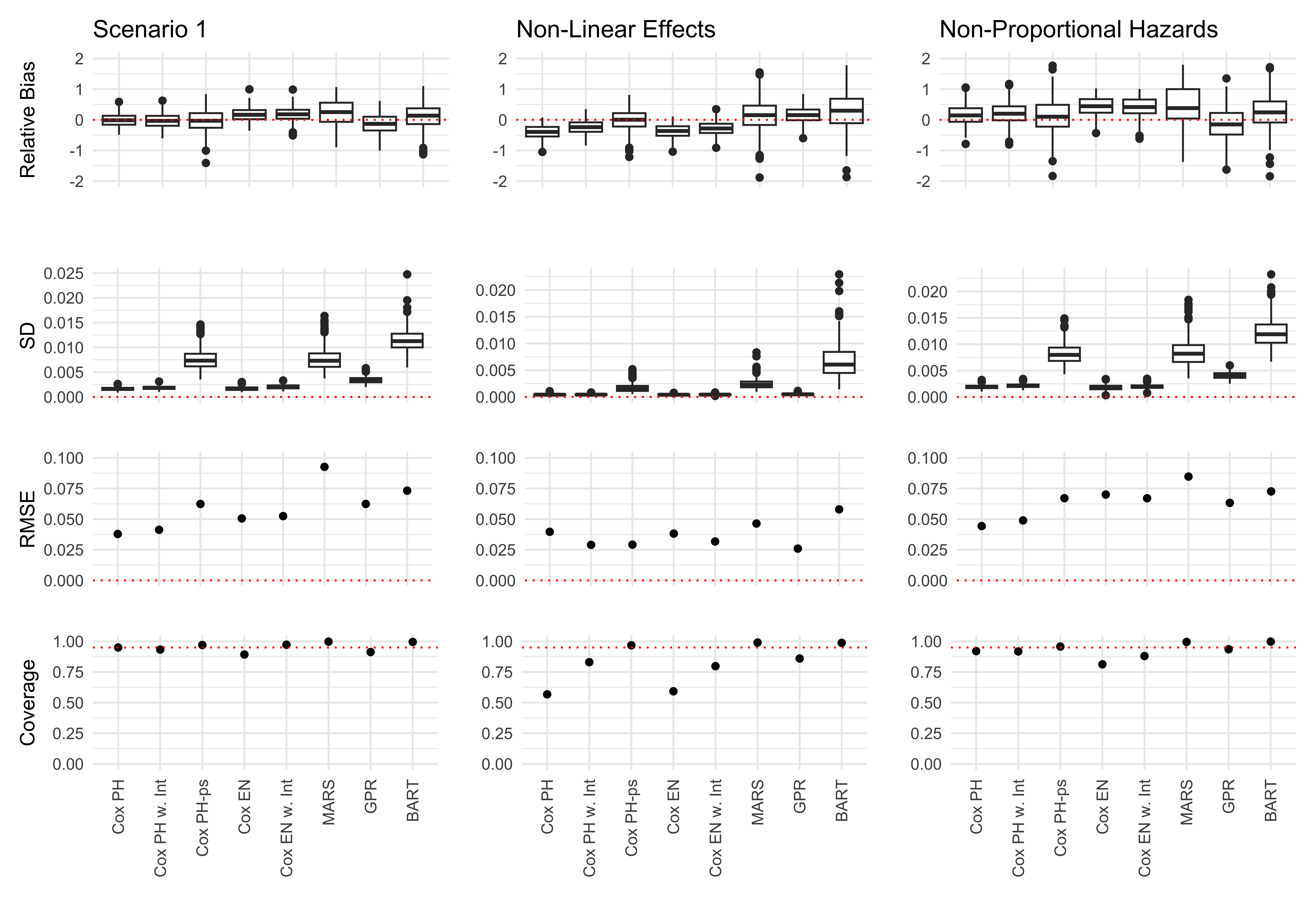}
\centering
\caption{Methods’ performances in estimating an environmental mixture’s effect on the survival probability difference scale.\label{fig4}}
\end{figure*}

In Figures \ref{fig1}-\ref{fig4} we show methods' performances in estimating the individual metal/metal mixture effect on both the hazard ratio and survival probability difference scales. Simulation performances for estimating the interaction effect can be found in the supplementary material. In Figures \ref{fig1} and \ref{fig2} we see performances in estimating the HR of an IQR change in the exposure to an individual metal (Figure \ref{fig1}) and overall metal mixture (Figure \ref{fig2}). All PH models performed relatively well for the base case scenario where the traditional Cox PH model is correctly specified. BART and MARS had the highest RMSEs due to their high variances in the estimated HR. MARS had particularly high variance, with the estimates and the standard deviations of the estimates for the metal mixture ranging from -11.6 to 0.97 and 0.4 to $2.7x10^{10}$, respectively. GPR performed slightly better than BART and MARS in terms of RMSE, however it has a lower coverage (0.76) when it comes to estimating the HR for the overall mixture effect. All methods had good coverage for estimating the individual metal in this scenario, ranging from Cox EN which had 92.0\% to BART which had 99.5\%. For the overall mixture effect, Cox EN had the next lowest coverage after GPR at 89\% while MARS had 100\% coverage.

For the scenario where the assumption of linearity is violated, RMSE increased across all methods and the coverage probability decreased for most methods. Due to the flexibility of MARS, the estimates had extremely high variance. The standard deviation of the estimated HR for an individual metal ranged from 0.35 to $3.7x10^{14}$, resulting in an RMSE of 84,989. BART also resulted in high variances in this scenario but was substantially less extreme, with an RMSE of 1.6.

We see an advantage in the discrete time survival analysis approaches in estimating HRs for the scenario where the proportional hazards assumption is violated. In this scenario, all methods had relatively similar RMSEs. However, MARS, GPR and BART all had much higher coverage probabilities (84.2\% - 99.5\%) compared to the PH approaches (5.0\% to 65.8\%). We can see in Figures \ref{fig1} and \ref{fig2} that GPR produced the lowest bias while maintaining low variance.

In Figures \ref{fig3} and \ref{fig4} we see performances in estimating the survival probability difference of an IQR change in the exposure to an individual metal (Figure \ref{fig3}) and metal mixture (Figure \ref{fig4}). For the difference scale, we found that while Cox PH-ps, MARS and BART have consistently higher variance compared to the other methods across scenarios, we do not observe as extreme values as what we did for the ratio scale, and all methods performed similarly. For scenario 1, all methods had low bias; thus, the more constrained methods (Cox PH and Cox EN with and without interactions) had lower RMSEs due to their lower variance. However, in scenario 2 when linearity is violated, the other four methods had less bias in their estimates of the survival difference. This results in similar RMSEs all around, with GPR yielding the lowest values (0.198 for individual metal and 0.582 for metal mixture). The three discrete time approaches and the Cox PH-ps model had higher coverage, 86.0\%-99.8\%, compared to the other methods, 56.8\%-83.0\%. When the hazards are not proportional (scenario 3), we see some similarities with the nonlinear case. Cox PH-ps, GPR, and BART appeared to have the lowest median relative bias. However, their higher variances result in higher RMSE compared to Cox PH and Cox EN with and without interactions. Additionally, there is less of a difference in coverage between the more flexible methods (91\% -99.8\%) and the more constrained methods (81.2\%-92\%).

\subsubsection{Graphical comparison of estimated and true curves}

\begin{figure*}
\includegraphics[scale=0.15]{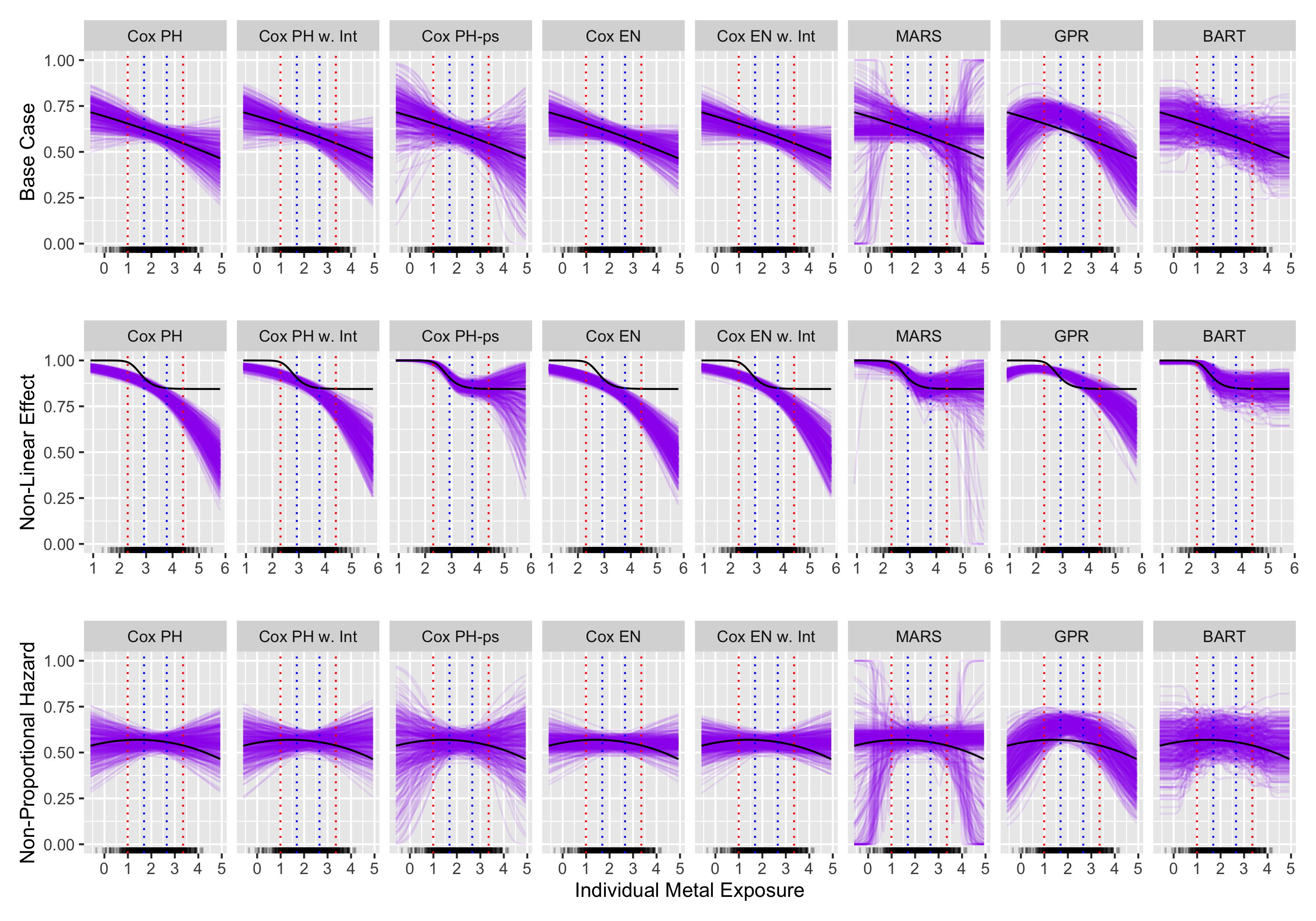}
\centering
\caption{Exposure-response curves in the form of survival curves over a range of an individual metal. We show the estimated curve for each of the 400 datasets simulated for each scenario. Black rug at bottom of plots shows density of metal from a randomly selected dataset. Blue dotted lines located at the 25th and 75th percentiles for the metal, indicating the values used for the IQR for the given metal. Red lines located at the 5th and 95th quantiles for the metal, where the majority of the data lies.\label{fig5}}
\end{figure*}

Figure \ref{fig5} shows the estimated survival curves across values of exposure of an individual metal for the 400 datasets simulated for each scenario. The MISE for the curves in Figure \ref{fig5} as well as the estimated survival curves for the nonlinear effects corresponding to other metals from the nonlinear scenario can be found in the supplementary material. Cox PH-ps and MARS did a decent job for the region where most of the data lie but had difficulty with the extremes of the exposure curve. BART appears to under-smooth. These three methods yielded large amounts of variability in their estimated survival curves. All other methods were more stable; however, lack the flexibility to capture the true response curve when nonlinearity occurs. Cox PH and Cox EN with and without interactions had the lowest MISE for both the base case (0.0008-0.0010 compared to 0.0017-0.0059) and the non-proportional hazards scenario (0.0009-0.0012 compared to 0.0018- 0.0067). However, for the nonlinear case, the discrete time survival analysis approaches and the Cox PH-ps model performed better (0.0004-0.0015 compared to 0.0025-0.0043).

\subsection{Results of Strong Heart Study Data Analysis}

We were interested in the association of the metal mixture on CVD in the SHS. Descriptive statistics for the baseline confounders and metal mixture components are presented in Table \ref{tab5}. Urinary concentrations of the metals are weakly to moderately correlated, Spearman correlation coefficients range from 0.01 to 0.26. The correlation matrix can be found in the supplementary material. We estimate the quantities of interest at the $80^{th}$ percentile of the observed follow-up time, which was 18.4 years. 904 participants (33.14\%) developed incident CVD.

\begin{table*}%
\caption{Baseline characteristics of Strong Heart Study participants (N=2728).\label{tab5}}%
\begin{tabular}{p{85mm} p{40mm} }
\hline
\textbf{Mean age (SE), years} & 56.1 (0.2) \\
\textbf{Female, \% } & 59.2 \\
\textbf{Education, \% } & \\
\hspace{3mm} No high school & 17.2 \\
\hspace{3mm} Some high school & 23.4\\
\hspace{3mm} Completed high school & 59.4\\
\textbf{Mean BMI (SE), kg/m2} & 30.4 (0.1) \\
\textbf{Smoking, \%} &  \\	
\hspace{3mm} Never &	29.3 \\
\hspace{3mm} Former & 32.9 \\
\hspace{3mm} Current	& 37.8 \\
\textbf{Mean Arsenobetaine (SE), $\mu$g/L} & -0.1 (0.02) \\
\textbf{Mean eGFR (SE), mL/min/1.73 m$^2$} & 97.5 (0.3) \\
\textbf{Median Creatine Adjusted Metal Concentrations (IQR), $\mu$g/g} & 	\\
\hspace{3mm} Arsenic*	& 8.42 (5.15, 14.32) \\
\hspace{3mm} Cadmium & 0.96 (0.62, 1.51) \\
\hspace{3mm} Molybdenum & 29.36 (20.46, 41.39) \\
\hspace{3mm} Selenium & 48.99 (36.73, 67.35) \\
\hspace{3mm} Tungsten & 0.12 (0.06, 0.23) \\
\hspace{3mm} Zinc & 561.30 (389.50, 805.35) \\
\hline
\multicolumn{2}{{p{125mm}}}{\footnotesize *Arsenic exposure is measured as the sum of inorganic and methylated arsenic species.}
\end{tabular}
\end{table*}

\begin{figure*}
\includegraphics[scale=0.15]{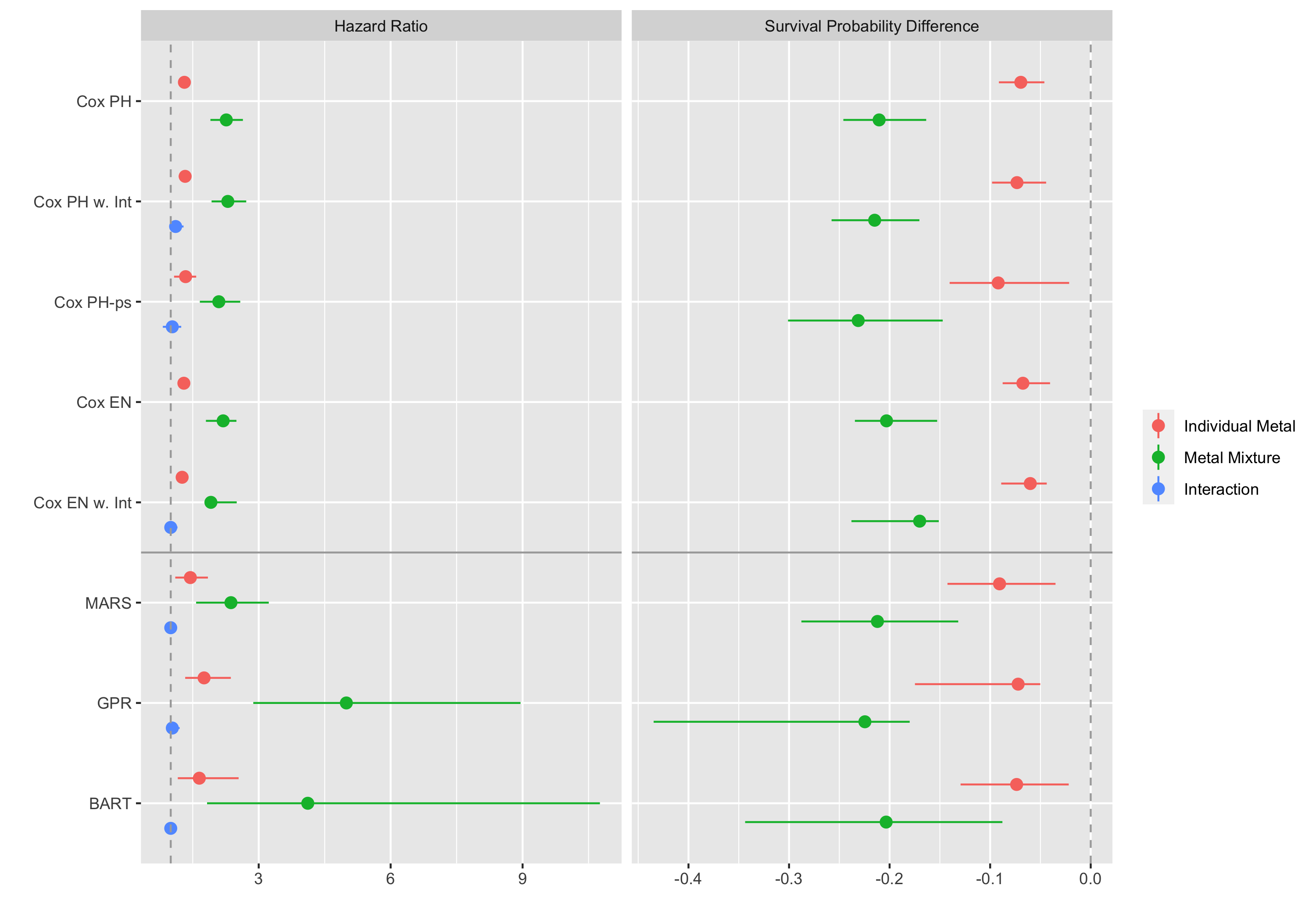}
\centering
\caption{Estimated effects on cardiovascular disease for the Strong Heart Study. Individual metal effect was estimated for Se. Metal mixture effect was estimated for the overall mixture comprised of As, Cd, Mo, Se, W, and Zn. Interaction effect was estimated between Se and W and are only shown for the models which allow for interaction estimations. Both the multiplicative (hazard ratio) and additive (survival probability difference) scale are shown.\label{fig6}}
\end{figure*}

We show the estimated effect of selenium and the overall metal mixture on CVD in the SHS. We chose to demonstrate the methods’ performance using selenium due to past evidence indicating an association of Se on cardiovascular disease.\cite{Zhao2022} We estimated the HR for an IQR change in Se and the overall metal mixture across the different modeling methods. For Se, the HR ranged for from 1.26 (1.18, 1.40), estimated via Cox EN with interactions, to 1.65 (1.16, 2.54), estimated via BART. For the metal mixture, the HR ranged from 1.91 (1.78, 2.50), estimated via Cox EN with interaction, to 4.12 (1.83, 10.75), estimated via BART. The estimated HRs for all methods can be seen in Figure \ref{fig6}. All methods found a significant, harmful (>1) effect of Se and the metal mixture on incident cardiovascular disease. However, the more flexible approaches found larger point estimates with larger confidence bands.

We additionally estimated the survival probability difference for an IQR change in Se and the metal mixture via the different modeling methods. The estimated survival probability difference for all methods can also be seen in Figure \ref{fig6}. For Se, the estimated survival probability difference ranged from -0.06 (-0.09, -0.04) to -0.09 (-0.14, -0.02). For the metal mixture, the estimated survival probability difference ranged from -0.17 (-0.24, -0.15) to -0.23 (-0.30, -0.15). The smallest estimated survival probability differences for Se and the metal mixture were estimated via the Cox EN with interactions and the largest estimated via the Cox PH-ps. All methods once again found a significant, harmful (<0) effect and the more flexible approaches found larger point estimates with larger confidence bands. Our finding are in line with what we saw in the simulation study, where the more flexible methods had wider confidence bands, potentially at the expense of reducing bias.

Figure \ref{fig6} also shows the point estimates and confidence intervals for the interactions between Se and W on the multiplicative scale. Cox PH and Cox EN do not have the capacity to detect an interaction, thus, they are not included. All other models find null effects, indicating that the effect of both Se and W at their $75^{th}$ percentile together is no different than the product of the effects of Se and W considered separately at their $25^{th}$ and $75^{th}$ percentile.

\section{Discussion}

We compared the performance of six methods in estimating the effect of an individual mixture component, overall effect of an environmental mixture, and interaction among mixture components on a time-to-event outcome. While the area of environmental mixture studies has received a great deal of attention, to our knowledge, limited consideration has been given to survival time outcomes.\cite{Joubert2022} We find that these types of research questions require special care, which might limit the use or alter the performance of commonly used methods for environmental mixture studies. We therefore offer a review of the currently available methods with extensions to the survival outcome setting, compare their performance across different scenarios via simulations, and apply them to a real-world scenario. 

In our simulation study we find that, in terms of flexibility in modeling exposure-response curves, the more constrained methods, such as Cox PH and Cox EN, tend to have lower variances compared to the more flexible methods. However, with less flexibility, more bias is introduced when estimating a specific effect. This bias appears to be more exaggerated when the proportional-hazard assumption is violated. Cox PH-ps, BART, GPR and MARS are more flexible and can better model departures from linearity and interactions. However, they tend to be more variable in their estimates and thus have lower power to detect a significant effect. There is a bigger advantage in the discrete time survival analysis approaches in estimating HRs for the scenario where the proportional hazards assumption is violated. In this scenario, all methods have relatively similar RMSEs. However, there may be cases where the discretization allowing for more flexible methods may be more important than retaining the continuous survival scenario due to the higher coverage probabilities associated with the discrete time survival approaches compared to the proportional hazards approaches. The preference between a constrained and a flexible method may vary depending on the degree of departure from linearity, the assumption of proportional hazards, or the effect size.

It is not possible to assess performance of the methods in all dimensions, we consider only selected accuracy methods. As a result, there are other strengths/weaknesses across methods which are not shown in detail in this study. For example, a common preference for methods for environmental mixture studies is variable selection. Of the methods we consider, only BART, EN and MARS can fully perform variable selection. We do not assess how well each method accurately performs variable selection. Variable selection and shrinkage methods mitigate the adverse impacts of multicollinearity commonly observed across environmental factors. Penalized regression such as EN is a good example of a shrinkage method. This may be a reason we find EN outperforms the standard Cox PH model in many ways. 

While variable selection is a preferable characteristic in an environmental mixture study for prediction, the inability to estimate valid post-selection inferences on models that use observed data for variable selection is a considerable limitation. For the frequentist methods, we estimate uncertainty of our estimates using the bootstrap because this is what is commonly done in practice.\cite{Domingo2022, Budtz2007, VanDerwerker2018} However, the nonparametric bootstrap may be invalid for these models. For example, the L1 norm penalized least square solution (otherwise known as the Lasso estimator) has been shown to be weakly consistent. These asymptotic results are particularly problematic when the true coefficient values are close to or exactly 0. In such cases, bootstrap sampling introduces a bias that remains asymptotically.\cite{Fu2000} The L1 component of EN’s penalized term implies their bootstrap confidence intervals are also invalid. Multiple approaches to post-selection inference have been proposed such as sample splitting, simultaneous inference, and conditional selective inference.\cite{Kuchibhotla2022} Incorporating these into our inference procedure may improve performance. Bayesian methods are more straightforward in that they directly use estimated posterior distribution to calculate uncertainty, thus efforts to improve Bayesian methods for environmental mixtures on survival outcomes may be beneficial. We also cannot be exhaustive of all possible real-world scenarios. For example, with the boom of big data, larger datasets with higher dimensional environmental mixtures are becoming increasingly more common. However, since we chose to have our simulated scenarios closely mimic the SHS, we restricted ourselves to demonstrating these methods’ performance under research questions similar to that of the SHS. This may limit the generalizability of our findings to other settings. 

An issue that may impact the performance of the discrete-time survival approaches is the specified number of bins into which the time variable is discretized into. While we would like to discretize time finely to preserve as much information as possible from the data, the larger the number of bins we choose the larger our augmented dataset becomes. Given the longer runtimes corresponding to the ML methods, the larger the dataset is the longer they will take to run, and it might not be feasible to consider a large number of bins. In practice, an assessment of how sensitive the effect estimates are to the number of bins used may be a worthwhile exercise to assess the impact of this choice. Additionally, censoring is a complex topic which arises when considering a survival outcome. For the simulation study, we set roughly ~70\% of observations as censored. This causes reduction in the effective sample size and can lead to a huge reduction in power. The more flexible methods often suffer from lack of power and require more data to provide more stable estimates. This may contribute to the high variability of some of the estimators, such as MARS. 

In the SHS, we found a significant negative effect of Se and the metal mixture across all methods. This is in line with previous findings, which have found Se to be associated with CVD outcomes, with no evidence of effect modification by other urinary metals.\cite{Zhao2022} Estimates across methods tended to be the same qualitatively, indicating the same direction of the effect, but different quantitatively, with magnitudes of the effect varying by method. Increased hazards/decreased survival probability were found at higher levels of metals, but the magnitudes varied. Although the overall conclusion is consistent, estimates may have different clinical impacts. The more flexible methods allowed for interactions and nonlinear effects of metals, but resulted in higher uncertainty. This reveals a substantial bias-variance tradeoff. Thus, to enhance reproducibility in environmental epidemiology, it is important to show whether results are robust across different modeling approaches for the same research question.

We conducted our real-world analyses using natural log transformed metal exposures. We also simulated our exposure variables to replicate the log transformed metal exposure from the SHS. Given the right skewed nature of environmental exposures, log transforming the exposures helps reduce this skewness which is often preferable for modeling methods to help satisfy certain assumptions or, in the nonparametric case, improve methods’ performance. While helpful, these transformations change the interpretations of our estimates by changing the scale of the exposures. The different scale can change our ability to detect true effects, which may become particularly challenging when we are dealing with interactions of two log transformed variables. While this is a common procedure for environmental mixtures analysis, it might lead to different conclusions, and it is unclear if it is beneficial for our scenario. However, conducting a simulation study to assess how different our results would be with and without variable transformations is beyond the scope of this paper.  

Previous overviews and comparisons of methods for estimating effects from an environmental mixture conclude that no one statistical method outperforms all others in every scenario and the choices of method should be selected based upon the scientific question of interest and/or any pre-analysis hypotheses.\cite{Taylor2016, Gibson2019, Hamra2018, Joubert2022, Sun2013} Others found little loss to using more flexible methods.\cite{Lazarevic2020} In this paper, we contribute to the growing area of research for environmental mixture analyses by extending past comparisons of modeling methods for analyzing the exposure to an environmental mixture to include the survival outcome setting. While ML methods performed well in some cases, their performances varied based on the scenario, suggesting they may be preferable for estimating the effect of an environmental mixture on a time-to-event outcome only under specific scenarios. Of the methods that we compared, none outperformed the others for all cases. In practice, various statistical methods should be applied given that one does not know the true data generating mechanisms, and it is helpful to see if findings are consistent using different methods. Many factors may influence a method’s performance, such as what the research question is, the sample size, follow-up time, correlations of metals, and nonlinearities/interactions. Importantly, and as demonstrated in this paper, the data type of the outcome of interest is also a large factor in impacting a method's performance. Our findings deviate from past results in many ways, and the results of this work underscore the need to develop or extend approaches to better estimate the effect of an environmental mixture specifically on a survival outcome.

\subsubsection*{Acknowledgements}

The Strong Heart Study was supported by grants from the National Heart, Lung, and Blood Institute contracts 75N92019D00027, 75N92019D00028, 75N92019D00029, and 75N92019D00030; previous grants R01HL090863, R01HL109315, R01HL109301, R01HL109284, R01HL109282, and R01HL109319; and cooperative agreements U01HL41642, U01HL41652, U01HL41654, U01HL65520, and U01HL65521; and by National Institute of Environmental Health Sciences grants R01ES021367, R01ES025216, R01ES032638, P42ES033719, and P30ES009089. We thank all the Strong Heart Study participants and Tribal Nations that made this research possible.

\bibliographystyle{ama}
\bibliography{main}  






\end{document}